\documentclass[prl,twocolumn,showpacs,amsmath,amssymb]{revtex4}
\usepackage[dvips]{graphicx}
\usepackage{bm}

\begin{document}

\title{Electron polarizability of
crystalline solids in quantizing magnetic fields \\
and topological gap numbers}

\author{Pavel St\v{r}eda$^{1}$, Thibaut Jonckheere$^{2}$ and
Thierry Martin$^{2,3}$}

\affiliation{$^{1}$Institute of Physics, Academy of Sciences of the
Czech Republic, Cukrovarnick\'{a} 10, 162 53 Praha, Czech Republic}
\affiliation{$^{2}$Centre de Physique Th\'{e}orique, Facult\'e de Luminy, Case 907, 13288 Marseille, France}
\affiliation{$^{3}$Universit\'e de la M\'editerran\'ee, Facult\'e de Luminy, Case 907, 13288 Marseille, France}

\date{\today}

\begin{abstract}
A theory of the static electron polarizability of crystals
whose energy spectrum is modified by quantizing magnetic  
fields is presented. 
The polarizability is
strongly affected by non-dissipative Hall currents induced by
the presence of crossed electric and magnetic fields: these can even
change its sign. 
Results are illustrated in detail for a two dimensional square lattice.
The polarizability and the Hall conductivity are respectively linked to 
the two topological quantum numbers entering the so--called Diophantine equation. 
These numbers could in principle be detected in actual experiments.  
\end{abstract}


\maketitle

The polarizability is usually presented
as an important property of insulators \cite{Ashcroft}.
An electric field imposed between the plates of
a capacitor penetrates into their interior, and charge
redistribution is induced within the system. This effect
is characterized by the polarizability:
the ratio of induced dipole moments per unit volume and
the local field. In metals this charge redistribution is
non-uniform because of the charge mobility and the local
electric field is non-uniform as well.
In the particular case that the
distribution of the atomic cores is not affected by the applied
electric field, it is the electronic charge redistribution only
which determines the polarizability of insulators as well as metals.

Crystalline solids have an energy spectrum composed of energy bands.
They become insulators whenever the Fermi energy
is located within an energy gap between these bands.
Additional energy gaps can be opened by applying 
strong
magnetic fields to two-dimensional (2D) as well as to three-dimensional
crystals \cite{Hofstadter,Koshino,Aoki}: 
the energy spectrum of 2D crystals is separated in an integer
number of subbands whenever the number of flux quanta contained
in the unit cell corresponds to a rational number \cite{Hofstadter}.
The number of electrons per subband is specified
by two topological gap numbers, one of which has been linked to the 
Hall conductance \cite{Thouless}. The present work  
establishes that for electrons in a ``strong'' periodic potential
subject to a magnetic field, the remaining gap number is directly linked 
to the polarizability.
We thus limit our attention to the static electron polarizability
and its dependence on the electron concentration, i.e. on the
Fermi energy $\mu$. This limitation 
allows to exclude dissipative processes from the consideration:
no current flow is allowed along the applied
electric field.  For the sake of simplicity, an ensemble of
spinless electrons at zero temperature is considered.
We will also limit our treatment to ``ideal'' crystals
with a rectangular unit cell of volume $a_z A_0$,
where $a_z$ is the lattice constant along the $\hat{z}$-direction
and $A_0 \equiv a_x a_y$ is the area in the $\hat{x} \hat{y}$ plane.

In the considered geometry,
the magnetic field is applied along the 
$\hat{z}$-direction, parallel with capacitor plates as well
as with the $z$-th crystallographic axis.
The external field due to the voltage drop between capacitor
plates is applied along the $\hat{y}$-direction.
The system is assumed to be open along the $\hat{x}$-direction,
(i.e. periodic boundary conditions) 
allowing non-dissipative current flow. This 
condition allows the direct comparison of the polarizability
with the topological gap numbers. 
Such condition could be realized in a Corbino geometry with 
cylindrical capacitor plates of large radius.

We start with the case of a vanishing magnetic field.
Generally, electrons are not equally distributed over
energy bands. The electron occupation $\tilde{s}_i^{(0)}(\mu)$ of 
the $i$-th band per volume $a_z A_0$ obeys the 
sum rule $a_z A_0 N(\mu)=\sum_{i} \tilde{s}_i^{(0)}(\mu)\equiv
\tilde{s}^{(0)}(\mu)$.
Here $N(\mu)$ denotes the electron concentration (the 
integrated density of states). Whenever the Fermi energy is located within
an energy gap, $\tilde{s}^{(0)}(\mu)=s$ ($s$ integer).
The wave functions of each band are  assumed to be of
Bloch-like form (i.e. extended) along the $\hat{x}$-direction while along
the $\hat{y}$-direction we consider a Wannier-like form (i.e. localized).
With this choice the mass-center positions
of the electrons along the $\hat{y}$-direction are well
defined by the expectation values of the $y$-coordinate.
Averaging over occupied states belonging to the $i$-th band
gives an average value
of the mass-center positions $\langle Y_i(\mu) \rangle$.
An electric field along the $\hat{y}$-direction, ${\cal{E}}_y$,
leads to a redistribution of the electron charge, which can be
characterized by the shifts of the mass-center positions
$\langle \Delta Y_i(\mu) \rangle$. These shifts are controlled
by the balance of the electric force  and of
the gradient force, due to the background crystalline
potential, which acts in opposite direction
to bring the electrons back into their equilibrium positions.
Within linear response in ${\cal{E}}_y$, the latter
is characterized by a force constant $K_i(\mu)$, so that:
\begin{equation} \label{Delta_Y_0}
- \, \tilde{s}_i^{(0)}(\mu) \, e {\cal{E}}_y \, - \,
K_i(\mu) \, \langle \Delta Y_i(\mu) \rangle \, = \, 0
\; .
\end{equation}
The resulting static electron polarizability $\alpha^{(0)}(\mu)$
is given as follows 
\begin{equation} \label{alpha_0}
\alpha^{(0)}(\mu) \, \equiv \, - \,
\frac{e {\sum_i} \langle \Delta Y_i(\mu) \rangle}{a_z A_0 \, 
{\cal{E}}_y} \, 
= \,
\frac{e^2}{a_z A_0} \, \sum_i 
\frac{\tilde{s}_i^{(0)}(\mu)}{K_i(\mu)}
\; .
\end{equation}
The above expressions are applicable even in the case
of a non-uniform electric field along the $\hat{y}$-direction
if the charge redistribution does not affect the values
of the force constants $K_i$. Strictly speaking, for
real metallic systems it might be a crude approximation.

Next, an external magnetic field splits the bands into subbands
\cite{Hofstadter,Koshino,Aoki}.
Because the geometry allows electrons to flow along
the $\hat{x}$-direction, the crossed electric ${\cal{E}}_y$
and magnetic $B$ fields give rise to a Hall current density $j_H$.
Consequently, an additional force -- the Lorentz force -- acts 
on the electron ensemble. The new balance condition along the
$\hat{y}$-direction becomes:
\begin{equation} \label{force_balance}
-  e N(\mu) \, {\cal{E}}_y -
\frac{B}{c} \, j_H - \frac{1}{a_z A_0}
\sum_i K_i(\mu) \, \langle \Delta Y_i(\mu) \rangle = 0
\, ,
\end{equation}
where the index $i$ now counts available sub-bands.
The Hall current density has the standard form
\begin{equation} \label{j_H}
j_H \, = \, - \, 
\frac{e^2}{h} \, \frac{\tilde{\sigma}(\mu)}{a_z} \, {\cal{E}}_y
\; ,
\end{equation}
where $\tilde{\sigma}(\mu)$ is dimensionless.
Introducing $\tilde{s}(\mu)$ to express the
mass-center shifts $\langle \Delta Y_i(\mu) \rangle$
in the form given by Eq.~(\ref{Delta_Y_0}), 
the force balance Eq. (\ref{force_balance}) becomes:
\begin{equation} \label{Diophantine_gen}
N(\mu) \, = \frac{\tilde{s}(\mu)}{a_z A_0} \, + 
\frac{\tilde{\sigma}(\mu)}{a_z \, 2 \pi l_B^2}
\; ,
\end{equation}
where $l_B=\sqrt{\hbar c/(e B)}$ is the magnetic length.

The effect of a strong magnetic field on Bloch electrons
is more pronounced in 2D systems, which motivates our choice of
a crystal formed by 2D planes, perpendicular to
the magnetic field, which are separated along the
$\hat{z}$-direction by $a_z$. The lattice constant $a_z$ is large enough that
electron transitions between planes are ruled out.
We thus consider  a single
layer only: a 2D crystal with electron concentration  $a_z N(\mu)$.
As was first noticed by Wannier \cite{Wannier} and recently
proved  by Kellendonk \cite{Kellendonk}, the necessary
condition for the appearance of an energy gap in such systems 
reads 
$N(\mu) \, = s/(a_zA_0) + \sigma/(a_z2 \pi l_B^2)$,
where $s$ and $\sigma$ are integers, often called  topological
gap numbers, and  $\sigma$ determines the value of
the quantum Hall conductance \cite{Thouless}. The comparison
of this so-called Diophantine equation with its general form
Eq.~(\ref{Diophantine_gen}) leads to the conclusion
that $s$ is related to the system polarizability.

The relation between polarizability and Hall
current is now illustrated 
with a tight-binding model on a square lattice,
$A_0=a^2$, assuming non-zero overlap between the nearest
neighbor sites only. At zero magnetic field it gives a single
energy band with cosine dispersion. 
The effect of the magnetic field is included with the Peierls 
substitution \cite{Hofstadter,Rammal}. 
This model was first used to obtain the ``Hofstadter butterfly'' energy
spectrum 
\cite{Hofstadter}. For rational
magnetic fields, satisfying:
\begin{equation}
\frac{A_0}{2 \pi l_B^2} = \frac{p}{q}
\; ,\quad  \quad
q= 2i+1
\end{equation}
(where $p$ and $i$ are integers), 
the energy spectrum
is composed of $q$ sub-bands well separated by energy gaps
for which topological gap numbers are uniquely defined
\cite{Thouless,Mouche}.

Choosing the Landau gauge $\vec{A} \equiv (-By, 0,0)$,
the Hamiltonian of the system remains periodic 
in the $\hat{x}$-direction. Zero-order wavefunctions
of each row of atomic orbitals
$\phi_a(x-am,y-an)$ along the $\hat{x}$-direction are thus
of the following standard form:
\begin{equation} \label{row_functions}
\Psi_{n,k_x}^{(0)}(\vec{r}) \, = \,
\sum_m e^{i k_x a m} \phi_a(x-am,y-an) 
\; ,
\end{equation}
where $k_x$ is the wave number. The vector
potential between atomic sites belonging to different rows
differs by $\Delta A_x = - B a \Delta n$, and the Peierls
substitution suggests to shift $k_x$ entering
the zero-field eigenenergies as follows:
\begin{equation} \label{Peierls_substitution}
k_x \, \rightarrow \,  
k_x \, - \, 2 \pi \, \frac{p}{q} \frac{n}{a}
\; ,
\end{equation}
so that phase factors in the overlap integrals
arise between orbitals of the neighboring rows.
Assuming zero-order eigenfunctions in the form of a linear
combination of row-eigenfunctions, Eq.~(\ref{row_functions}),
the coefficients $c_n(k_x)$ have to satisfy the following
Harper's equation \cite{Harper,Rammal}: 
\begin{eqnarray} \label{Harper_eq}
c_n(k_x) \left[
2 \cos \left( k_x a - 2 \pi \frac{n p}{q} \right) +
\frac{E-E_a}{\Delta V} \right] +
\nonumber \\
c_{n-1}(k_x) + c_{n+1}(k_x) = 0
\, ,
\end{eqnarray}
where $E_a$ is the energy of the atomic orbitals (chosen to be zero)
and $\Delta V$ is the overlap strength (the zero field band width
is  $8\Delta V$).
The modulus of the $c_n$ coefficients are periodic with a period
$q$, but their amplitude differs by a phase  $\beta$:
$c_{n+q} = e^{i \beta} c_n$ \cite{remark_1}.
For a given $\beta \in [0,2 \pi]$,
Eq.~(\ref{Harper_eq}) gives a system of $q$ equations,
and the $c_{n}(k_x,\beta)$ are eigenvectors of a $q \times q$ matrix.
The eigenvalues $E_{\beta}(k_x)$ give the $q$ sub-bands, 
each of them composed of energy branches determined by the phase
$\beta$, which are periodic in $k_x$ with period $2 \pi/a$.
Because of the periodicity of the $|c_n(k_x,\beta)|$,
it is natural to define Wannier-like functions,
which are extended (localized) along the $\hat{x}$
($\hat{y}$) direction:
\begin{equation} \label{Wannier_functions}
w_{n,k_x,\beta}(\vec{r}) \, = \,
\sum_{\lambda=-i}^{+i} c_{n+\lambda}(k_x,\beta) \,
\Psi_{n+\lambda,k_x}^{(0)}(\vec{r})
\; .
\end{equation}
The average mass-center position in the $\hat{y}$-direction
reads:
\begin{equation} \label{Y}
Y_{n,\beta}(k_x) \equiv
\int y |w_{n,\beta,k_x} (\vec{r})|^2  d{\bf r} =
a n + Y_{\beta}^{(B)}(k_x)
\, ,
\end{equation}
where
\begin{equation} \label{delta_Y_B}
Y_{\beta}^{(B)}(k_x) \, = \,
a \sum_{\lambda=-i}^{+i} \lambda |c_{\lambda}(k_x,\beta)|^2
\; ,
\end{equation}
which vanishes at zero magnetic
field. 

\begin{figure}[h]
\includegraphics[angle=0,width=3.3 in]{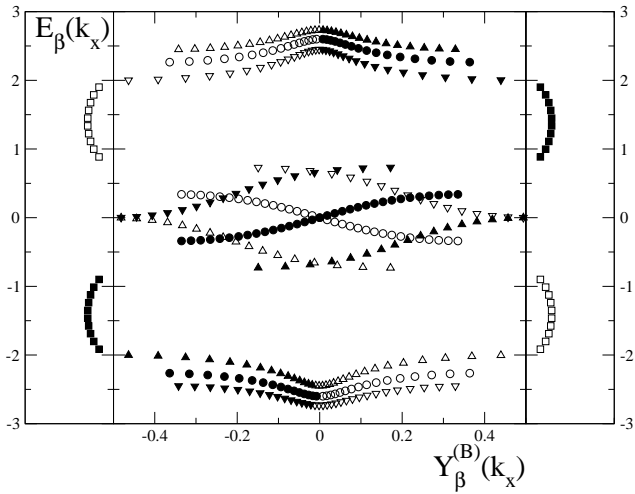}
\caption{Energy spectrum $E_{\beta}(k_x)$ (in units of the overlap strength $\Delta V$) 
as a function of the center of mass coordinate 
$Y_{\beta}^{(B)}(k_x)$ (in units of the lattice constant) for the fraction $p/q=1/3$, shown for
$\beta =0$ (down triangles), $\beta = \pi/2$ (circles) and $\beta = \pi$ (up triangles).
Empty (filled) symbols correspond to positive (negative) velocity along the $x$ axis. The left and
right sides show the magnetic edge states  inside the energy gaps.}
\label{fig:mc}
\end{figure}
The eigenvalue problem is mathematically 
equivalent to the one found for 
a weak periodic potential in a strong magnetic field~\cite{Streda07}, and leads to the same spectrum
as a function of $k_x$, except that the values of $p$ and $q$ are interchanged.
However, the spectrum presented as a function of the mass center position
differs substantially from that of 
the weak periodic potential.
 For the ratio $p/q=1/3$,
the energy spectrum is composed of  three
subbands, each of them formed by the branches obtained for all values of $\beta$. 
The spectrum is symmetric with respect to zero energy 
because of electron-hole symmetry.
This is illustrated in Fig. \ref{fig:mc}, where only three values 
$\beta=0,\pi/2$ and $\pi$ are shown.  
The branches $\beta=0,\;\pi$ correspond to the subband edges,
while $\beta=\pi/2$ characterizes the central branch.
At the crystal edges, determined by the values $n_L$ and 
$n_R$ of the row index $n$, the natural condition that
$c_{n_L-1} = c_{n_R+1}=0$ leads to the appearance of edge states
formed by contributions from each of the energy branches $\beta$.
Two types of edge states appear, non-magnetic edge states and 
magnetic ones~\cite{Rammal,Streda94} which are responsible
for the quantum Hall effect. The latter have opposite velocities at opposite edges,
and are shown in Fig. \ref{fig:mc} for $p/q=1/3$.
Note that in the thermodynamic limit, 
the mass-center positions of edge states
can be identified with the edges of the physical system.

Next, we switch on ${\cal{E}}_y$, which
shifts the position of the atomic orbitals.
The force trying to return electrons back to their
original positions is assumed to be linear in this shift,
with a proportionality (``force'') constant $K\equiv m_0 \Omega_0^2$,
with a confinement frequency
$\Omega_0$ ($m_0$ is the free electron mass). Within 
linear response with respect to ${\cal{E}}_y$ the 
mass-center positions of atomic orbitals 
are shifted along the $\hat{y}$-direction
by the distance $-e {\cal{E}}_y/(m_0 \Omega_0^2)$ from their
equilibrium positions $n a$.
This shift enters the Peierls substitution, 
Eq.~(\ref{Peierls_substitution}), giving rise to an
additional shift of $k_x \rightarrow k_x + \Delta k_x$: 
\begin{equation} \label{delta_k}
\Delta k_x \equiv
\frac{\omega_c^2}{\Omega_0^2} \, \frac{e {\cal{E}}_y}{\hbar \omega_c}
\, ,
\end{equation}
where $\omega_c=eB/m_0 c$ is the cyclotron frequency. 
The resulting shift of the mass-center position reads:
\begin{equation} \label{Delta_Y_beta}
\Delta Y_{\beta}(k_x) \, = \, - \,
\frac{e {\cal{E}}_y}{m_0 \Omega_0^2}
\left( 1 - l_B^{-2} \, \frac{d Y_{\beta}^{(B)}(k_x)}{d k_x} \right)
\, .
\end{equation}
The electric field ${\cal{E}}_y$ also gives rise to a potential energy,
and up to the lowest order the eigenenergies are modified simply by an
additive contribution $e {\cal{E}}_y Y_{n,\beta}(k_x)$. As a result the
expectation value of the velocity along the 
$\hat{x}$-direction changes as: 
\begin{equation} \label{delta_v_x}
\Delta v_x(\beta,k_x) \, = \, \frac{e {\cal{E}}_y}{\hbar} \,
\frac{d Y_{\beta}^{(B)}(k_x)}{d k_x}
\, .
\end{equation}
Consequently, a non-zero current density along $\hat{x}$-direction
(the Hall current) is induced by ${\cal{E}}_y$. It can be expressed
in the form of Eq.~(\ref{j_H}) with $\tilde{\sigma}(\mu)$
given as: 
\begin{equation} \label{sigma}
\tilde{\sigma}(\mu) =
N_{\beta}^{-1} \sum_{\beta=1}^{N_{\beta}} \int d k_x \;
f_0 \left( E_{\beta}(k_x) - \mu \right) \,
\frac{d Y_{\beta}(k_x)}{d k_x} 
 ,
\end{equation}
where $f_0(E-\mu)$ denotes Fermi-Dirac distribution function
and $N_{\beta}$ denotes number of branches.
When the Fermi energy $\mu$ lies within the energy gaps,
$\tilde{\sigma}$ approaches an integer value $\sigma$, which
can differ from zero due to the presence of magnetic edge states. 

\begin{figure}[h]
\includegraphics[angle=0,width=3.3 in]{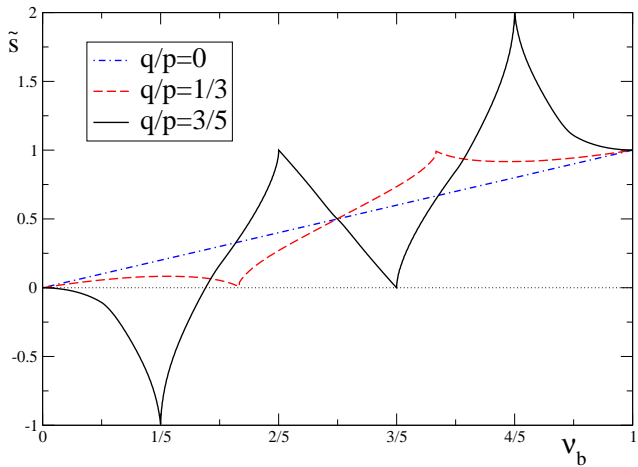}
\caption{Dependence of the effective topological number
$\tilde{s}$,
on the band filling factor $\nu_b$ for the ratio
$p/q=0, \,1/3$ and $3/5$.
}
\label{fig:s}
\end{figure}

Using the Eq.~(\ref{Delta_Y_beta}) defining the average shift of the 
mass-center positions and the identity Eq.~(\ref{Diophantine_gen}),
the electron polarizability can be written as:
\begin{equation} \label{alpha}
\alpha(\mu) =
\frac{e^2}{m_0 \Omega_0^2}
\left( N(\mu)  - \frac{\tilde{\sigma}(\mu)}{a_z 2 \pi l_B^2} \right)
 = \frac{e^2}{m_0 \Omega_0^2} \, \frac{\tilde{s}(\mu)}{a_z A_0}
\, .
\end{equation}
For the tight-binding model considered here, the dependence 
of $\alpha(\mu)$ is fully specified by the effective topological
number $\tilde{s}(\mu)$. The latter is plotted  
as a function of the filling factor of the tight-binding band,
$\nu_b \equiv a_z A_0 N(\mu)$ in Fig.~\ref{fig:s}
for several values of the ratio $p/q$.
The case $p/q=0$ corresponds to the zero magnetic field, for which
$\tilde{s} \rightarrow \tilde{s}^{(0)} = \nu_b$ as follows from
Eq.~(\ref{Diophantine_gen}). For non-zero ratios a rich  behavior
is induced by the applied magnetic field. Whenever the Fermi
energy is located within the energy gap between subbands, i.e. if 
$\nu_b$ is given by integer multiples of $1/q$, the
effective topological number $\tilde{s}(\mu)$ takes an integer
value which is just equal to the topological gap number.
As seen in Fig. \ref{fig:s}, for $p/q = 1/3$, the total contribution of the lower and the upper 
sub-bands to the polarizability vanishes.  
In this particular case the
applied electric
force acting on carriers belonging to these sub-bands is fully
compensated by the Lorentz force due to the induced non-dissipative
edge Hall currents.
For the case $p/q = 3/5$ the magnetic field even changes the
sign of $\tilde{s}$ and consequently the sign of the polarizability.
At half-filling of the band ($\nu_b=1/2$), the Hall current 
vanishes and $\tilde{s}$ approaches its zero-field value $1/2$.
The central symmetry of all curves around this half-filling point
($\nu_b=1/2$, $\tilde{s}=1/2$) is a consequence of
the electron-hole symmetry of the single tight-binding band.

In this letter we have analyzed the interplay between the Hall
current and the electron polarizability for the tight-binding
model of a crystalline solid. This interplay is controlled 
by the compensation of the external electric force with the 
two other forces acting on the electrons:
the gradient force, due to the periodic 
background potential, which is related to the polarizability,
and on the other hand the Lorentz force, which determines the
non-dissipative Hall current.
The presence of the quantizing
magnetic field induces a rich and complex behavior for the electron
polarizability, which can even change sign as a function
of the electron concentration.

Most importantly, we provide an answer to the long-standing question:
{\it what measurable quantity is determined by the topological gap
number $s$ entering the Diophantine equation}? While it has  been understood for
a long time that $\sigma$ determines the quantum Hall effect,
a similar interpretation of $s$ has so far remained unclear.
We have shown that $s$ is directly linked to the static electron polarizability.
Contrary to the quantum Hall effect, the proportionality
constant between the polarizability and the topological number 
$s$ is a material dependent quantity rather than an universal constant, 
which can even depend on the magnetic field. 
The independence of the force constant $m_0 \Omega_0^2$ on the
magnetic field considered here is a mere consequence of the use of the Peierls substitution
which is justified when $\Omega_0\gg \omega_c$.
A shrinking of atomic orbitals by the
magnetic field, which leads to a suppression of the overlap strength 
and thus to a decrease of the overall band width, 
could be included by the substitution
$\Omega_0^2 \rightarrow \Omega_0^2 + \omega_c^2/4$. 

The predicted effect requires that a few flux quanta penetrate the 
lattice area $A_0$. Even for solids with a lattice constant of the order 
of $10$\AA, this requires exceedingly large fields ($\sim 10^3$ T). If, however, the 
magnetic field is tilted  with respect to the high--symmetry crystallographic 
directions (here $\hat{z}$), an effective area
(which is larger than $A_0$) can accomodate a few flux quanta, so that the above 
condition can be reached with experimentally available fields \cite{Aoki}.
This condition can also be reached in 2D arrays of quantum dots or 
antidots with a lattice constant $\sim 100$ nm. However, such (anti) dots need 
to be weakly coupled in order to reproduce a tight binding energy 
spectrum. Their size reduction (which implies a weak overlap and well separated 
atomic states) is thus a necessary
condition for observing the predicted effect, and it still represents an 
experimental challenge.  

The authors acknowledge support from
Grant No. GACR 202/05/0365 of the
Czech Republic. P.S. acknowledges support from the
Institutional Research Plan No. AV0Z10100521,
and thanks CPT (UMR6207 of CNRS) for its hospitality.


\begin{thebibliography}{99}
\bibitem{remark_1}
If periodic boundary conditions were applied along 
the $\hat{y}$-direction, the standard wave number $k_y$ would be introduced 
instead of the branch index $\beta$.
\end{thebibliography}
\end{document}